\documentclass[runningheads,a4paper]{llncs}

\usepackage{cmap}
\usepackage[utf8]{inputenc}
\usepackage[T1]{fontenc}
\usepackage{ mathrsfs }
\usepackage{float}

\usepackage{graphicx}

\usepackage{subcaption}
\usepackage{multirow}
\usepackage{booktabs}
\usepackage{adjustbox}
\usepackage{array}
\newcolumntype{R}[2]{%
    >{\adjustbox{angle=#1,lap=\width-(#2)}\bgroup}%
    l%
    <{\egroup}%
}
\newcommand*\rot{\multicolumn{1}{R{45}{1em}}}

\usepackage[ngerman,english]{babel}
\addto\extrasenglish{\languageshorthands{ngerman}\useshorthands{"}}

\usepackage[%
rm={oldstyle=false,proportional=true},%
sf={oldstyle=false,proportional=true},%
tt={oldstyle=false,proportional=true,variable=true},%
qt=false%
]{cfr-lm}
\usepackage[math]{blindtext}

\usepackage{cite}

\usepackage{paralist}

\usepackage{csquotes}

\usepackage{microtype}

\usepackage{url}
\makeatletter
\g@addto@macro{\UrlBreaks}{\UrlOrds}
\makeatother

\usepackage{xcolor}

\usepackage{pdfcomment}
\hypersetup{hidelinks,
   colorlinks=true,
   allcolors=black,
   pdfstartview=Fit,
   breaklinks=true}
\usepackage[all]{hypcap}

\usepackage[capitalise,nameinlink]{cleveref}
\crefname{section}{Sect.}{Sect.}
\Crefname{section}{Section}{Sections}

\usepackage{xspace}

\DeclareFontFamily{U}{MnSymbolC}{}
\DeclareSymbolFont{MnSyC}{U}{MnSymbolC}{m}{n}
\DeclareFontShape{U}{MnSymbolC}{m}{n}{
    <-6>  MnSymbolC5
   <6-7>  MnSymbolC6
   <7-8>  MnSymbolC7
   <8-9>  MnSymbolC8
   <9-10> MnSymbolC9
  <10-12> MnSymbolC10
  <12->   MnSymbolC12%
}{}
\DeclareMathSymbol{\powerset}{\mathord}{MnSyC}{180}

\usepackage{tikz}
\newcommand{\chapternote}[1]{{%
  \let\thempfn\relax
  \footnotetext[0]{\emph{#1}}
}}

\let\llncssubparagraph\subparagraph
\let\subparagraph\paragraph
\let\subparagraph\llncssubparagraph

\begin{document}

\title{Like trainer, like bot? Inheritance of bias in algorithmic content moderation}


\author{Reuben Binns\inst{1}, Michael Veale\inst{2}, Max Van Kleek\inst{1}, Nigel Shadbolt\inst{1}}

\institute{
Department of Computer Science, University of Oxford\\
\email{reuben.binns|max.van.kleek|nigel.shadbolt@cs.ox.ac.uk}\and
Department of Science, Technology, Engineering and Public Policy (STEaPP), University College London\\
\email{m.veale@ucl.ac.uk}
}

\maketitle

\begin{abstract}
The internet has become a central medium through which `networked publics' express their opinions and engage in debate. Offensive comments and personal attacks can inhibit participation in these spaces. Automated content moderation aims to overcome this problem using machine learning classifiers trained on large corpora of texts manually annotated for offence. While such systems could help encourage more civil debate, they must navigate inherently normatively contestable boundaries, and are subject to the idiosyncratic norms of the human raters who provide the training data. An important objective for platforms implementing such measures might be to ensure that they are not unduly biased towards or against particular norms of offence. This paper provides some exploratory methods by which the normative biases of algorithmic content moderation systems can be measured, by way of a case study using an existing dataset of comments labelled for offence. We train classifiers on comments labelled by different demographic subsets (men and women) to understand how differences in conceptions of offence between these groups might affect the performance of the resulting models on various test sets. We conclude by discussing some of the ethical choices facing the implementers of algorithmic moderation systems, given various desired levels of diversity of viewpoints amongst discussion participants.

\end{abstract}


\section{Introduction}
\chapternote{Accepted to the Proceedings of the 9th International Conference on Social Informatics (SocInfo 2017), Oxford, UK, 13--15 September 2017 (forthcoming in Springer Lecture Notes in Computer Science).}
Online platforms, as `curators of public discourse'~\cite{gillespie2010politics} or digital extensions of the public sphere~\cite{dahlberg2001internet}, have become important spaces for opinion and debate. While social media, news websites, and question--answer forums enable exchange of diverse viewpoints~\cite{halpern2013social}, aggressive, offensive or bullying comments can stifle debate, drive people away, and lead to intervention by regulators or law enforcement. Yet over-restrictive moderation can similarly send users elsewhere. Consequently platforms' terms of use, content policies and enforcement measures often attempt to bound acceptable discourse~\cite{ksiazek2015civil}.
 
Moderation by employees, contractors, users or volunteers is an explicitly human endeavour. However quantity of content makes manually and rapidly vetting each item very costly, driving interest in automated content classification. Automatic detection used to require curated blacklists of banned words, but as these are difficult to maintain as language, norms, and gaming strategies change, more novel means involve training machine learning algorithms on large corpora of texts manually annotated for aggression, offence or abuse. According to the description of Google's `Perspective API'~\cite{perspectiveapi}, platforms might wish to predict the `impact a comment might have on a conversation', giving `realtime feedback to commenters or help moderators do their job'. Microblogging platform Twitter~\cite{wagner2017} and comment plug-in Disqus~\cite{rohrer2017} are pursuing similar efforts.
 
While automating content moderation might lighten staff and volunteer burden, its norms hinge on raters' judgements within training data. Where multiple implicit or explicit communities exist --- particularly where participation in labelling is not balanced --- this might penalise content exhibiting particular views or vernacular. The global imposition of raters' norms might affect diversity and participation on the platform.

This paper explores methods for detecting of potential bias in algorithmic content moderation systems. We experiment with a series of text classifiers using an existing dataset of 100,000 Wikipedia comments manually scored for `toxicity' (annotators' questions, Figure \ref{table:ratingq}). To examine how differences between norms of offence might result in different classifications, we built different classifiers from demographically distinct subsets of the population responsible for labelling the training data. We focus on gender as a demographic variable potentially associated with differences in judgements of offence, primarily due to the ease of drawing balanced samples compared to other available variables (age, education). We do \textit{not} intend to establish generalisable conclusions about gender and offence

\section{Background and related work}
We do not attempt to define aggression, offence or harassment in this paper (for an overview of definitions, see~\cite{wolak2007does}). Suffice to say, different logics for automated and semi-automated moderation exist, including the promotion of `quality' comments online~\cite{Diakopoulos:2016a}, the flagging of hate speech or bullying~\cite{gagliardone2015countering,schrock2011problematic,tokunaga2010following,burnap2016us}, or the maintenance of imagined `networked publics' that `allow people to gather for social, cultural, and civic purposes'~\cite{boyd2012}. 

Content moderation can have real and lasting effects on the direction of and participation in conversations in the digital public sphere. Norms of acceptability do not exist in a vacuum, as they are reinforced by prior standards, and they are also malleable. Previous research on online comments has found that by intervening in certain ways, news organizations can affect the deliberative behavior of commenters~\cite{stroud2015changing}, altering the kinds of comments made (e.g. thoughtful or thoughtless)~\cite{sukumaran2011normative}, and users' perceptions of the content they comment on~\cite{anderson2014nasty}.

Systems for automated content moderation began with primarily manually encoded rules and features, later becoming more inductively driven. Early work focussed on identifying abusive and hostile messages or `flames' with manually crafted features (such as evaluated regular expressions or word lists) and decision trees~\cite{spertus1997smokey}. These systems gave way to more general machine learning--based inferences, with bag-of-words and topic modelling both popular approaches. While newer machine learning techniques have been applied lately (see~\cite{Pavlopoulos2017}), many relevant issues such as contextualisation are a bottleneck more widely across machine learning research~\cite{schmidt2017survey}.

As censorship and free speech are issues at the heart of democratic politics~\cite{mill1999liberty,feinberg1985offense}, inductive systems that seek to automatically reduce the visibility of certain contributions are unlikely to escape the scrutiny of those worried about `algorithmic bias'. What (if anything) should be filtered is and has always been a matter for heated societal debate. Content considered `abusive' by some might to others be partisan disagreement. Studying news platform comment moderation, Diakopoulos and Naaman found that media organisations acknowledge that moderators bring their own biases to the evaluation of standards~\cite{diakopoulos2011towards}.

Recent years have seen a rapidly intensifying focus on the way that bias enters computational systems, linked especially to the consequences of systems that `profile' individuals and make decisions that relate to their lives. In these fields, efforts to understand and mitigate illegal bias or general unfairness in areas from loan acceptance to word embeddings have propagated~\cite{feldman2015certifying,Calders1:2012,BolukbasiCZSK16a,IslamBN16}. Yet algorithmic content moderation has some distinctions from the current main trajectories of `discrimination-aware data mining' (DADM). In particular, while DADM attempts to ensure fairness across individuals that share characteristics protected by anti-discrimination legislation, such as race, gender, religion, pregnancy, or disability, issues in algorithmic content moderation are not always of this type. While there might be instances where protected groups are directly affected --- the filtering of African American Vernacular English, for example --- practical issues seem more likely to relate to creating diverse, welcoming (and often legally compliant) places to be online. Here, some determined equitable distribution of \emph{viewpoints} might be of more interest than representation of protected groups \textit{per se}, although it is nonetheless likely that individuals within certain demographics share some norms of offence, gender being one example often studied~\cite{johnson1985sex,sutton2001bitches,jay1992cursing}. 

In applied contexts, firms might be more interested in user-bases, political ideologies, or other platform-specific divides than protected characteristics --- for example, avoiding classifiers that more often flag `liberal' rather than `conservative' comments. We might say this system is `unfair' to some viewpoints.  In the majority of cases, while it is unlikely these groups will be well-defined or self-declared, there may be practical methods for platform operators to segment users for analytic purposes. We focus in this paper on exploratory methods over defined groups, using gender as an illustrative example, but with the explicit caveat that gender will rarely be the prime grouping of interest.

\section{Pragmatic approaches for exploring biases by altering test and training sets}

Our general question is: how do latent norms and biases affect the operation of offence detection systems? Specifically, do the norms of offence held by people who contributed the training data result in classifiers which systematically favour certain norms of offence over others? 

The usual way to evaluate a classifier H is to define a loss function L, which measures the extent to which its predictions \textit{\^{Y}} approximate the ground truth of the phenomena of interest \textit{Y}. Normally, there is only one version of the ground truth, i.e. one set of labels for \textit{Y}. In this case, we want to measure biases between classifiers relative to different \textit{norms of offence}, which are characterised by sets of labels applied to a corpus of comments \textit{C} in natural language $[c_1, c_2, ..., c_n]$ which score each comment along some axis, (e.g. $0 = \textrm{`not offensive'}$, $1 = \textrm{`offensive'}$). If we take sets of ground truth we believe correspond to different norms, then applying chosen loss functions to the predictions \textit{\^{Y}} and these ground truths help us better explore the nature of its `bias' towards or against certain norms. There are other times where it is possible to alter the training data, but less easy to alter the test data. While the training dataset can always be split, this says little about performance where the domain is different --- for example, where the classifier will be deployed on a website where comments have much less metadata. If the training data contain characteristics believed to be correlated with norms, a collection of classifiers $[H_1, H_2, ..., H_m]$ corresponding to \textit{m} norms can be trained. Instead of evaluating a single classifier against several test sets, several classifiers (for example, labelled by those differing in inferred political standpoints) can be compared to a single, domain-relevant ground truth.

To demonstrate this approach, we examined an existing dataset of offence labels including demographic information about labellers. Our aim was to find a demographic variable likely to be associated with differences in attitudes about what texts are considered offensive. We chose to use gender because it was an easily understood, accessible demographic attribute of the labelling population, which enabled us to select large and equally sized sub-populations for new training sets, which (we hypothesised) would differ in their definitions of offence.

\section{Data Sources and Methodology}

\begin{figure}[tb]
\begin{minipage}[b]{.5\textwidth}
\scriptsize
\centering
\begin{tabular}{p{0.9\textwidth}}
\toprule
\textbf{Rate the toxicity of this comment}\\\midrule
- \textit{Very toxic} (a very hateful, aggressive or disrespectful comment that is very likely to make you leave a discussion)\\
- \textit{Toxic} (a rude, disrespectful or unreasonable comment that is somewhat likely to make you leave a discussion)\\
- \textit{Neither}\\
- \textit{Healthy contribution} (a reasonable, civil or polite contributions that is somewhat likely to make you want to continue a discussion)\\
- \textit{Very healthy contribution} (a very polite, thoughtful or helpful contribution that is very likely to make you want to continue a discussion)\\\bottomrule 
\end{tabular}
\subcaption{Question provided to raters}\label{table:ratingq}
\end{minipage}
\begin{minipage}[b]{.5\textwidth}
\centering
\input{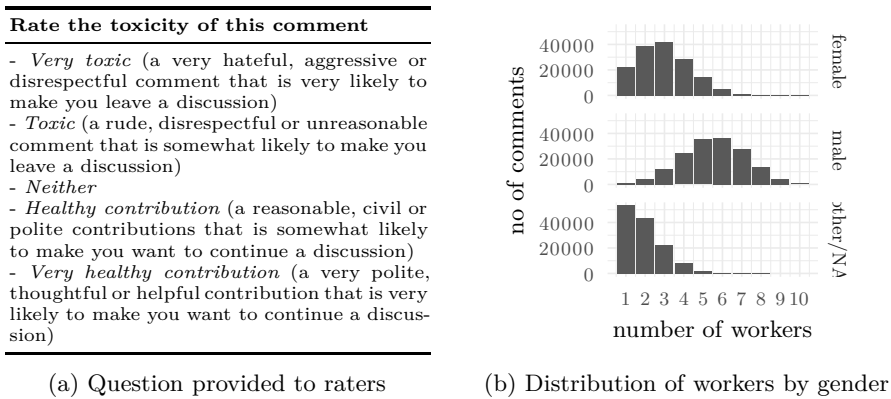}
\subcaption{Distribution of workers by gender}\label{fig:genderraters}
\end{minipage}
\caption{Dataset statistics and rater question.}
\end{figure}

We trained classifiers with an existing dataset from the Wikipedia Detox project (used in~\cite{wulczyn2017ex}). It features 100,000 annotations of Wikipedia talk page comments manually labelled by workers on the Crowdflower platform. Each comment is labelled by 10 workers for `toxicity' (see  Figure \ref{table:ratingq}). Workers optionally provide demographic data, including age group, gender (restricted to male, female, other) and educational level. Noticeably, workers are not evenly distributed across reported gender (28.6\% female, 55.6\% male, 15.8\% unreported/other), and there are fewer females per comment (see distribution in Figure \ref{fig:genderraters}). Comments without both male and female raters were excluded, as were raters not providing gender or the few selecting `other'.

Considering the dataset before training classifiers, we measured agreement on toxicity within male and female annotators with Krippendorff's alpha (due to amenability to missing data, see~\cite{hayes2007answering}))\footnote{95\% bootstrapped CIs (500 replicates) were calculated with R's \emph{rel} package~\cite{kriprel}.}, and average measures of how toxicity scores per comment differs between male and female annotators.

We then trained multiple text classifiers on various subsets of the original training data. Our model building process involved converting comments into \textit{ngrams} (ranging from 1-2) at a maximum of 10,000 features, constructing a matrix of token counts, applying a TF-IDF vectoriser, before training a logistic regression model.\footnote{Code and data available at \url{https://github.com/sociam/liketrainer}.} We first trained a classifier using all of the original training data (as used in the original study by~\cite{wulczyn2017ex}), to benchmark our modelling process against prior work and ensure that reasonable performance could be expected of classifiers built. We found that our models achieved an AUC score of 0.914 (the highest-performing classifier in~\cite{wulczyn2017ex} achieved a score of 0.96). We then used a bootstrapping method to sample new subsets of annotators in order to build various classifiers from these subsets. For each comment which had both male and female raters, we selected 10 male/female annotators at random with replacement. We then took the average score for these 10 sampled raters. 30 sets of training data were generated this way: 10 male, 10 female, and 10 a balanced mix. These  data were used to train 30 different offensive text classifiers as above. Each set of classifiers was tested against unseen `male', `female' and `balanced' rated test data, sampled in the same way as above.

\section{Results}
We found evidence of difference between male and female annotators regarding the labelling of comments as `toxic'. Firstly, inter-rater agreement (Krippendorff's alpha) was significantly lower for women (.468 [.457, .478 (95\% bootstrapped CI)]) than for men (.494 [.484, .503 (95\% bootstrapped CI)]); female annotators were less likely to agree with each other's offence scores than males. Furthermore, on average female annotators found comments less `toxic' than male counterparts. By comment, average female toxicity scores were 0.043 [-0.048, -0.038 (95\% CI)] lower than male ones.

\begin{figure}[tb]
\input{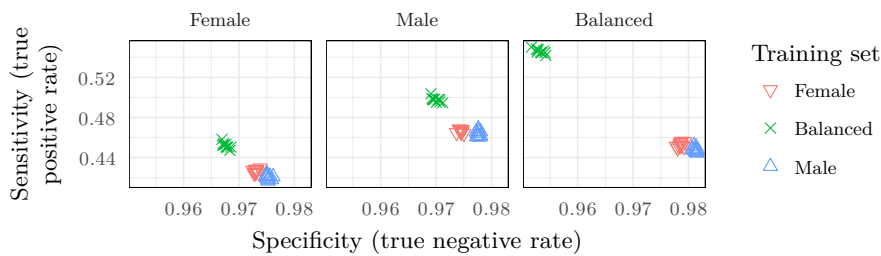}
\caption{Sensitivity and specificity, by training set (points) and test set (facets).}
\label{fig:sensspec}
\end{figure}

Having established some gender differences in norms of offence, we proceeded to analyse if differences were also distinguishable in classifier performance on various test datasets. We found that both classifiers trained on male \textit{and} those trained on female annotated data are less sensitive to female-labelled test data than to male-labelled test data (see Figure \ref{fig:sensspec} and Table \ref{table:ratingq}). True positive rates for female-labelled test data were 0.42 (`male' classifiers) and 0.43 (`female' classifiers), while true positive rates for male-labelled test data were higher at 0.46 (`male' classifiers) and 0.47 (`female' classifiers). We did not find such disparities in terms of specificity; `male' and `female' classifiers had similar true negative rates when tested on male and female-labelled data. In terms of the bias/fairness definition given above, both types of classifier could be considered `unfair' to women, insofar as they exhibit more false positives when attempting to replicate \textit{women's} collective judgments than \textit{men's}. In other words, speech that female annotators collectively \textit{did not} find offensive was more likely to be mis-classified as such by both `male' and `female' classifiers. We found that mixed-gender classifiers had higher sensitivity across all three test sets.

\begin{figure}[h!]
\input{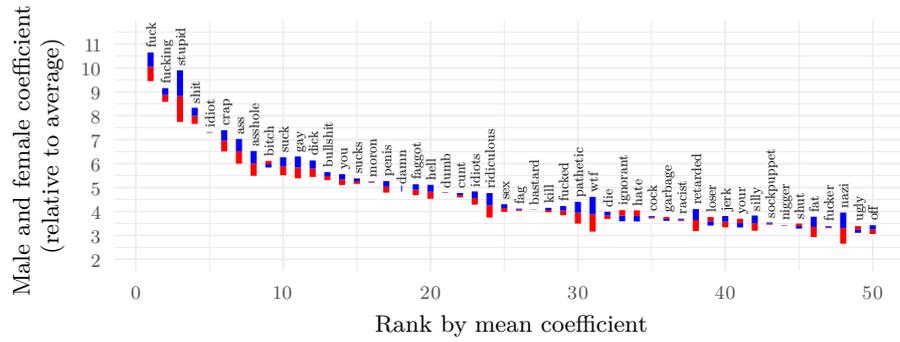}
\caption{Highest 50 average model coefficients greater than 2 by gender, ranked by average. Ends of blue and red lines indicate male and female coefficients.}
\label{fig:offplot}
\end{figure}

We also compared coefficients of the `male' and `female' classifiers. We took features used by the classifiers and calculated average coefficients across the 10 classifiers created for each gender. Selecting ngrams most strongly associated with offensive classifications (with a coefficient value of more than 2) for both `male' and `female' classifiers, we found significant overlap between `male' and `female' classifiers. The most offensive terms for `male' classifiers however tended to be more strongly associated with offence than the same terms for `female' classifiers. Few terms were indicated to be more offensive to `female' than `male' classifiers, and for those that were, the margin was small (see Figure \ref{fig:offplot}). 

\section{Discussion}
Online communities exhibit norms regarding acceptable speech. Who gets to define these norms is a contentious matter. It might be within the remit of company executives, or undertaken in consultation with users and other stakeholders. Large platforms often have teams dedicated to identifying and removing offensive content, while also relying on users flagging content, both as means of detection and as `rhetorical justification' for censorship~\cite{crawford2016flag}. Moderation privileges may not be universal --- volunteers may be self-appointed (e.g.~Reddit), appointed through semi-democratic processes (e.g.~Wikipedia), or implicitly through reputation (e.g.~StackExchange). Whoever defines them, norms of acceptability are rarely static, consistent or uncontested. Discussion forum Reddit features many sub-fora exhibiting different norms, leading to frequent arguments between users, staff and executives across sub-fora over what kinds of posts should be allowed~\cite{centivany2016values}. What counts as acceptable is therefore always a subjective matter.

It seems inevitable that introducing inductive automated content moderation systems risks amplifying these subjective norms and potentially exacerbating conflicts around them. These concerns may be compounded in cases where training data is decontextualised from the domain of application --- increasingly common as more sophisticated classifiers, or classifiers for new or nascent communities utilise multiple external sources of data. The Perspective API, partly borne out of the Wikipedia Detox project, is now being used to moderate comments on the New York Times website~\cite{bassey2017}. One approach attempting to consider community differences is explored in~\cite{chandrasekharan2017bag}. Even if training data is only taken from where the moderation is occurring, it might introduce and reproduce historical biases or patterns incompatible with the changing nature of community norms. Where community standards are internally contested or in flux, automated content moderation will likely compound pre-existing platform conflicts.

In so far as the bounds of acceptable discourse are inherently contestable, we argue (relatively uncontroversially) that there can be no such thing as a `neutral’ classifier. Even in politically homogeneous environments with broad agreement about the offensiveness of all training data, an automated system would still constrain future speech on the basis of what has been deemed acceptable/unacceptable in the past. A healthy public sphere must also be capable of evolving, sometimes prohibiting speech which was previously acceptable and at other times relaxing prohibitions as social mores change.

At the same time, there can be no formula determining the extent to which different viewpoints need to be reflected in order for a classifier to be deemed fair. While anti-discrimination law might give anchoring to DADM methods, we cannot expect anything nearly so readily formalisable in this domain. Stakeholder norms differ --- Some communities (or groups within them) might legitimately desire homogeneous `safe spaces’ in which otherwise offensive/inoffensive speech is permitted or prohibited; others might want to positively promote more diversity in discussion and thus aim to create classifiers which strike a balance between error rates which optimises diverse participation. In the latter case, one might aim to either minimise the false positive rates (making the conversation more permissive), or minimise the false negatives (less permissive), depending on whether subscribers to the under-represented norm are driven away primarily by over-zealous censorship or by exposure to comments they deem offensive.

While we use gender, as a common variable, it will often be that either protected characteristics are not the core concern, or that traits of concern are unavailable. Focussing on the training and test sets according to performance metrics of interest to find problematic patterns is a practical first step for platforms today given existing data, skillsets and methodologies. Going beyond gender, unsupervised methods such as clustering or manually identifying users by behaviour might help identify groups with conflicting views of offense. Finally, it is important to recognise that more civil discourse --- if that is indeed a desirable goal --- is unlikely to be achieved solely through moderation, whether manual or algorithmic; it also requires careful consideration of community dynamics, interface design, and rationales for participation.

\section{Concluding remarks}
This case study aimed to illustrate methods and metrics for exploring bias in text classification tasks where learned concepts are inherently contestable, and to prompt reflection on the range of ethical considerations that should be taken into account by designers of algorithmic moderation systems, and the platforms that deploy them. While we do not conclude that automated moderation systems are `sexist' (and did not seek to show this), we demonstrated how particular training sets may be biased in ways that are worth investigating prior to implementing such systems. Defining fairness in these systems strictly is likely impractical: these are highly complex, changing and contested concepts, and even were definitions arrived at, it would be unlikely that regular platforms held sufficient data on commenters and raters to operationalise them. Instead we illustrate an exploratory approach involving varying the test and training sets, which we believe to be a useful first step for organisations looking to implement automated content moderation, or test, monitor or evaluate technologies they are using.

As algorithmic content moderation approaches become more pervasive,  platforms deploying them will face difficult choices with significant implications for the development of community discussions and digital public spheres. If individuals with certain viewpoints feel unwelcome, then polarisation online will likely continue to increase. People are drawn to online communities in part due to discursive norms and editorial policies, but algorithmic enforcement of those norms and policies could warp them in unforeseen ways. Platforms would therefore be advised to introduce such systems only with careful consideration and ongoing measurement; we hope that the methods discussed here can help.

\subsubsection*{Acknowledgments}
Authors at the University of Oxford were supported under \emph{SOCIAM: The Theory and Practice of Social Machines}, funded by the UK Engineering and Physical Sciences Research Council (EPSRC) under grant number EP/J017728/2. Michael Veale was supported by EPSRC grant number EP/M507970/1. The UCL Legion High Performance Computing Facility (Legion@UCL) supported part of the analysis. Thanks additionally go to three anonymous reviewers for their helpful comments.

\bibliographystyle{splncs03}
\bibliography{refs}

All links were last followed on June 14, 2017.
\clearpage
\section*{Appendix}

\begin{table}[h!]
\centering
\begin{subtable}{.4\textwidth}
\begin{tabular}{crlll}
&& \multicolumn{3}{c}{\textit{Test set}}\\
\parbox[t]{4mm}{\multirow{4}{*}{\rotatebox[origin=c]{90}{\textit{Training set}}}} && \rot{Female} & \rot{Male} & \rot{Balanced}\\\cline{3-5}
& \multicolumn{1}{r|}{Female} & 0.427 & 0.467 & 0.453\\
& \multicolumn{1}{r|}{Male} & 0.420 & 0.464 & 0.448\\
& \multicolumn{1}{r|}{Balanced} & 0.452 & 0.498 & 0.546 \\
\end{tabular}
\caption{Sensitivity (true +ve rate)}
\end{subtable}
\begin{subtable}{.4\textwidth}
\begin{tabular}{rlll}
& \multicolumn{3}{c}{\textit{Test set}}\\
& \rot{Female} & \rot{Male} & \rot{Balanced}\\\cline{2-4}
\multicolumn{1}{r|}{Female} & 0.973 & 0.974 & 0.979\\
\multicolumn{1}{r|}{Male} & 0.975 & 0.978 & 0.981\\
\multicolumn{1}{r|}{Balanced} & 0.968 & 0.970 & 0.953 \\
\end{tabular}
\caption{Specificity (true -ve rate)}
\end{subtable}
\caption{Average performance by demographic of training and test sets.}
\end{table}

\end{document}